\documentclass[10pt,letterpaper]{article}
\usepackage[top=0.85in,left=2.75in,footskip=0.75in]{geometry}

\usepackage{amsmath,amssymb}

\usepackage{changepage}

\usepackage{textcomp,marvosym}

\usepackage{cite}

\usepackage{nameref,hyperref}

\usepackage[right]{lineno}

\usepackage[nopatch=eqnum]{microtype}
\DisableLigatures[f]{encoding = *, family = * }

\usepackage[table]{xcolor}

\usepackage{array}

\usepackage{tipa}

\usepackage{longtable} 
\usepackage{pdflscape} 
\usepackage{booktabs}
\usepackage{adjustbox}
\usepackage{arydshln}
\usepackage[numbers]{natbib}
\usepackage{listings}
\usepackage{xcolor}
\usepackage{enumitem}

\usepackage{cleveref}
\newcounter{suppsec}
\renewcommand{\thesuppsec}{S\arabic{suppsec}}
\crefname{suppsec}{Appendix}{Appendices}
\Crefname{suppsec}{Appendix}{Appendices}

\newcommand{\suppappendix}[1]{%
  \refstepcounter{suppsec}%
  \paragraph*{\thesuppsec\ Appendix: #1}%
}

\newcolumntype{L}[1]{>{\raggedright\let\newline\\\arraybackslash\hspace{0pt}}p{#1}}
\newcolumntype{C}[1]{>{\centering\let\newline\\\arraybackslash\hspace{0pt}}p{#1}}

\lstdefinestyle{pythonstyle}{
    language=Python,           
    backgroundcolor=\color{gray!20}, 
    basicstyle=\ttfamily\footnotesize, 
    frame=single,              
    breaklines=true,           
    tabsize=4,                 
    rulecolor=\color{black},   
    showstringspaces=false,    
    keywordstyle=\color{blue}, 
    stringstyle=\color{red},   
    commentstyle=\color{green!50!black}, 
}

\newcolumntype{+}{!{\vrule width 2pt}}

\newlength\savedwidth

\raggedright
\usepackage[aboveskip=1pt,labelfont=bf,labelsep=period,justification=raggedright,singlelinecheck=off]{caption}

\makeatletter
\renewcommand{\@biblabel}[1]{\quad#1.}
\makeatother

\usepackage{lastpage,fancyhdr,graphicx}
\usepackage{epstopdf}
\fancyheadoffset[L]{2.25in}
\fancyfootoffset[L]{2.25in}
\newcommand{\eg}{e.\,g.\,}
\newcommand{\ie}{i.\,e.\,}

\begin{document}
\vspace*{0.2in}

\begin{flushleft}
{\Large
\textbf\newline{Towards clinical adoption of voice and speech as measures of health: the need for harmonization} 
}
\newline
\\
Nicholas Cummins\textsuperscript{1,2\Yinyang, *},
Vikram Ramanarayanan\textsuperscript{3,4\Yinyang},
Daniel Low \textsuperscript{5,6\Yinyang},
Fabio Catania\textsuperscript{7\Yinyang},
Si-Ioi Ng\textsuperscript{8,9\Yinyang},
Caterina Botelho\textsuperscript{10,11},
Judith Dineley\textsuperscript{1},
Abir Elbeji \textsuperscript{12},
Julie M. Liss\textsuperscript{9},
Paula Andrea Pérez-Toro \textsuperscript{13,14},
Visar Berisha \textsuperscript{9\ddag},
Thomas Quatieri\textsuperscript{15\ddag}
\\
\bigskip
\textbf{1} Institute of Psychiatry, Psychology and Neuroscience, King's College London, UK
\\
\textbf{2} NIHR Biomedical Research Centre: Maudsley, UK
\\
\textbf{3} Modality.AI Inc, San Francisco, California, USA
\\
\textbf{4} University of California San Francisco, San Francisco, California, USA
\\
\textbf{5} Child Mind Institute, New York, NY, USA
\\
\textbf{6} Department of Psychology, Harvard University, Cambridge, USA
\\
\textbf{7} McGovern Institute Massachusetts Institute of Technology (MIT) Cambridge, USA
\\
\textbf{8} Department of Language Science and Technology, The Hong Kong Polytechnic University, Hong Kong SAR of China 
\\
\textbf{9} Arizona State University, Tempe, AZ, USA
\\
\textbf{10} INESC-ID, Lisbon, Portugal; 
\\
\textbf{11} Sword Health, Portugal
\\
\textbf{12} Department of Precision Health, Luxembourg Institute of Health, Strassen, Luxembourg
\\
\textbf{13} Pattern Recognition Lab, Friedrich-Alexander-Universitat Erlangen-Nurnberg (FAU), Erlangen, Germany
\\
\textbf{14} Chair for AI in Healthcare and Medicine, Technical University of Munich (TUM), Munich, Germany
\\
\textbf{15} Lincoln Laboratory, Massachusetts Institute of Technology, Lexington, MA, USA
\\
\bigskip

%
%
\Yinyang These authors contributed equally to this work as joint first authors.
\ddag These authors also contributed equally to this work as joint senior authors.

* nick.cummins@kcl.ac.uk

\end{flushleft}
\section*{Abstract}
Speech and voice are multidimensional signals that capture both communicative intent and underlying physiological processes, providing a unique, non-invasive window into health. Analyzing these signals has the potential to yield digital biomarkers that (i) provide scalable, objective measurement tools for research and clinical care and (ii) reflect the presence or progression of diverse conditions, including neurological, psychiatric, respiratory, and cardiovascular disorders. Realizing this promise, however, requires the field to overcome pervasive reproducibility and generalizability issues due to heterogeneous data collection, processing, and analysis practices. A major source of this heterogeneity is how underlying acoustic measures themselves are defined and computed. In this paper, we outline key considerations across the speech biomarker discovery lifecycle, from data collection through machine learning modeling to clinical interpretation, needed to achieve reliable, reproducible, and clinically translatable results. Chief among these is the need for harmonization efforts to start from common, precisely specified measure definitions. As a first step, we therefore provide definitions, physiological correlates, and computational implementations for a minimal, clinically interpretable set of core speech measures spanning respiration, phonation, articulation, and fluency. We close by discussing ongoing standardization efforts and the open challenges that remain in advancing the adoption of speech- and voice-based digital biomarkers.



\section*{Introduction}
\label{sec:intro}
Beyond conveying explicit linguistic messages, speech and voice signals encode substantial implicit physiological, emotional, and cognitive information, capturing a unique, multifaceted view of the speaker’s health. This encoding reflects the complexity of speech production, intertwining cognition and a finely coordinated interplay of respiratory, phonatory, and articulatory functions. For these reasons, speech production is one of the most complex and finely coordinated forms of human behaviour.

Historically, voice and speech have served as implicit markers of health, with spoken language as a primary channel for clinician–patient interactions. The emergence of ambient scribing systems has enabled the routine capture and documentation of spoken interactions in clinical settings, highlighting speech as a rich and scalable source of information \cite{Dineley_tech_matters, lawrence2026use, stanton2025evaluating}.  Beyond its linguistic content, each utterance reflects a complex combination of physiological, cognitive, and motor processes. Clinicians implicitly leverage cues such as voice quality, fluency, and lexical richness for diagnostic insights, underscoring voice and speech as a vital, yet underutilized, data source in healthcare.

A surge in research on objective voice and speech analysis~\cite{idrisoglu2023applied} suggests substantial potential for health outcome monitoring through automated computational methods that quantify measurable changes in speech characteristics. Continuing advancements in signal processing and machine learning  demonstrate that specific alterations in speech—modulated by health conditions—can be objectively assessed and monitored \cite{ng2026end, ramanarayanan2022speech, milling2022speech, low2020automatedreviewspeech, cummins2015review}. Additionally, the proliferation of remote data collection technologies, such as smartphone and web-based applications, enables frequent and scalable speech assessments in daily life at a cadence far exceeding what is feasible in clinical settings. Collectively, these advances create the conditions for more ecologically valid, frequent voice and speech assessment within emerging models of virtual and personalized healthcare.

To realize this vision, it is critical that we begin to address the methodological variability concerns that currently hamper the field. Voice and speech processing research for health is currently grappling with reproducibility issues. Diverse and underreported approaches to speech data collection, processing, and analysis complicate reproducibility and cross-study comparisons are critical barriers to scaling speech-based health assessments and bridging the gap between research and clinical adoption~\cite{ramanarayanan2022speech}. 

Harmonizing these elements would involve establishing consistent variables, measure-extraction procedures, and study-design protocols across research groups and applications. It also includes adopting software-engineering practices that enable reproducibility and embracing open-science principles for transparent dissemination of data, code, and methods. This would strengthen reproducibility, enable systematic comparison across studies, and, crucially, help build a standardized evidence base for the clinical integration of automated voice and speech analysis.

A universal, one-size-fits-all approach to automated clinical voice and speech analysis is inherently challenging given the diverse clinical and research contexts in which speech is used. Voice and speech characteristics vary greatly across individuals, languages, contexts, and measurement methodologies, while clinical conditions introduce additional layers of complexity. These sources of variability mean that any harmonization framework must strike a balance between standardization and flexibility to remain clinically meaningful and broadly applicable. To address these challenges, this work brings together a broad set of domain experts and outlines a structured process for harmonizing clinical voice and speech analysis.

Before we dive deeper into this process, it is important to justify and define the key terms used in this paper. In the speech-analytics health literature, terminology is used inconsistently, with overlapping terms applied to similar concepts across disciplines. In this context, we distinguish voice (phonation) from speech: voice refers to the acoustic signal generated by the larynx, whereas speech denotes the realized form of spoken language shaped by articulatory and prosodic processes.  Additionally, we use the term \textit{speech measure}; defined as a quantitatively derived value intended to measure an underlying attribute or construct (e.g., speech motor control, prosodic variability), generally specified together with its intended interpretation and evidence supporting that interpretation. This aligns with influential validity frameworks in psychological measurement that treat validity as concerning the meaning and interpretation of scores, not a property of the instrument alone, and that frame scale/test development as the creation of a valid measure of an underlying construct~\cite{cronbach1955construct,messick1995validity,american1985standards}. 

We have adopted the term \textit{speech measure} as the least discipline-loaded term for a mixed clinical–engineering readership: \textit{feature} typically implies an engineered model input, while \textit{metric} is easily confusable with model-evaluation metrics. \textit{Biomarker} is another commonly used term in the literature, the use of this term carries stronger claims about clinical meaning. It generally refer to objective biological indications (i.e., substance, structure, or process) that can be accurately and reproducibly measured from inside or outside the patient \cite{ramanarayanan2022speech}. Relatedly, a recent consensus has arisen regarding the use of the term \textit{vocal biomarker}, it states that this term should only be adopted once a measure has undergone extensive validation procedures within a defined context of use \cite{Pizzimenti2026}. Our use of \textit{measure} fits within this consensus framework \cite{Pizzimenti2026}, as it succinctly and clearly conveys the message that \textit{this value is intended to measure something} while keeping the evidentiary burden explicit via construct-interpretation validity.

Returning to our harmonization aim, we propose a framework to support the systematic use of clinically interpretable acoustic measures, with primary emphasis on the definition and standardization of the features themselves. Specifically, we introduce a core set of  \emph{clinically interpretable} measures spanning respiration, phonation, articulation, spectral structure, and fluency, selected for their established links to motor, respiratory, cognitive, and affective constructs. We provide precise definitions alongside normative values for American English, derived from healthy speakers on standardized tasks, to enable consistent interpretation and cross-study comparability. Using this framework, we highlight that the validity of these measures is contingent on their alignment with elicitation tasks and clinical constructs; without this alignment, the measures themselves are not valid. To support transparent and reproducible application, we also release open-source code for feature extraction with standardized settings.


\section{Project Lifecycle in Voice and Speech Research for Health}
\begin{figure}[p]
     \centering \includegraphics[width=1\linewidth]{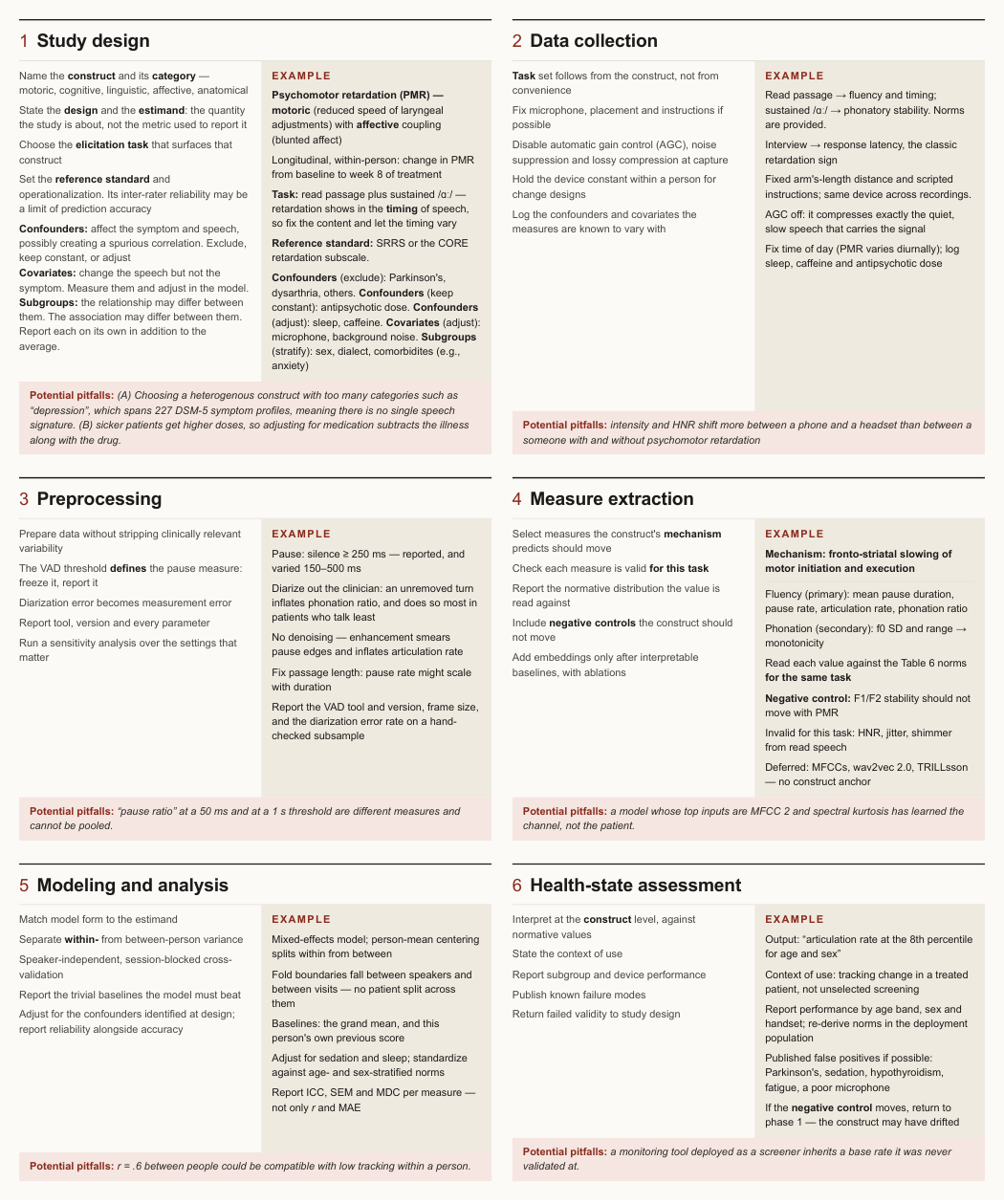}   
    \caption{Project Lifecycle in Voice and Speech Research  for Health. 
    }
    \label{fig:lifecycle}
\end{figure}

The automated analysis of voice and speech has the potential to yield digital biomarkers associated with a range of health outcomes~\cite{ramanarayanan2022speech, Pizzimenti2026, ng2026end}.
Realizing this potential requires research efforts that support the transition from speech measures to vocal biomarkers by building broad, replicable evidence bases, grounding claims in well-defined clinical constructs, and validating them through robust measurement protocols \cite{alonso2024definitions, Pizzimenti2026}. This requires establishing clear, consistently reproducible associations with the physiological or pathological states being measured (e.g., how the tone of voice reflects vocal-fold oscillation), supported by harmonized frameworks that ensure reliability, accuracy, and comparability across studies.  The lifecycle of a voice and speech data project comprises a series of structured phases that are essential for generating replicable and generalizable clinical evidence (\Cref{fig:lifecycle}). This framework spans from the initial project design to the final interpretation and application of results.

The lifecycle begins with \emph{study design}, where researchers define the clinical constructs of interest, select specific tasks that elicit them, and specify inclusion criteria, covariates, and evaluation measures. Details on study design factors and speech elicitation consideration are provided in \Cref{subsec:StudyDesign} and \Cref{subsec:elecit} respectively. 

In the next phase, \emph{data collection}, factors such as microphone type, placement, recording instructions, and digitization parameters (e.g., sampling rate, bit depth, and compression) may influence the acoustic signal, particularly in remote or bring-your-own-device settings. Additional contextual factors, including time of recording and environment, may also affect signal quality and warrant consideration; see \Cref{subsec:DataCollect} for further details. 

The `raw' recordings can then undergo \emph{preprocessing}, where the aim is to prepare data for analysis without stripping away clinically relevant variability. Example steps may include segmentation, diarization, or denoising. Further details on issues relating to preprocessing are give in \Cref{subsec:preprocess}. 

The critical step in the lifecycle is \emph{feature extraction}, in which speech is transformed into quantifiable measures that index the constructs defined at study design.  Variability at this stage—stemming from software toolkits, parameter choices, or inconsistent definitions; \eg \cite{cummins2025protocol, dineley2024variability}. Combined with upstream variability introduced during data collection poses one of the greatest barrier to reproducibility and motivates our focus on harmonization. These concerns are covered in more detail in \Cref{subsec:featex}.

Subsequent phases include \emph{modeling and analysis}, where statistical or machine learning methods functionally map measures to health outcomes (\Cref{subsec:HealthAssessment}), and \emph{health-state assessment and deployment}, where measures must prove reliable, interpretable, and clinically useful under real-world conditions (\Cref{subsec:DataAnalysis}). 

\subsection*{Spotlight on feature extraction}
Methodological heterogeneity in automated voice and speech analysis stems from differences in design choices such as elicitation tasks, recording conditions, and feature extraction. This heterogeneity remains a primary barrier to building replicable evidence bases. Moreover, sources of variability across the full signal acquisition and analysis pipeline can interact with one another, making it difficult to compare findings across studies or aggregate results into clinically meaningful conclusions. Addressing this challenge requires harmonization at the level of measurement, guided by a clear principle: measures should be \emph{clinically anchored}, \emph{task-appropriate}, and \emph{comparable}. Accordingly, this work centers on the definition and harmonization of clinically interpretable acoustic measures.

Elicitation protocols determine which aspects of voice and speech can be observed, and therefore which clinical constructs can be reliably measured. For example, sustained phonation reveals phonatory stability through measures such as Cepstral Peak Prominence (CPP); rapid syllable repetition tasks help expose articulatory control, while reading passages can be used to measure differences in fluency through pause rates and articulation speed. Conversely, a sustained phonation task offers no insight into speech timing measures, and phonation measures can be less reliable when estimated from free speech samples due to the influence of articulation. Measures that are anchored to constructs through their elicitation tasks ensure that results remain interpretable and reliable, even with limited sample sizes. This step also enables aggregation across studies and supports the development of a cumulative evidence base for clinical adoption. 

A further source of variability in measures arises from differences in recording conditions and devices, which influence extracted measures and affect comparability across studies~\cite{cummins2025protocol, maryn2017mobile, jannetts2019assessing, solera2021transfer, botelho2022challenges, dineley2023towards, liu2024clever}. Signal intensity-based measures, for example, depend on device characteristics and speaker–microphone distance, and should therefore be interpreted with caution in bring-your-own-device studies. While preprocessing tools can help minimise noise in analytical pipelines, their excessive use should be avoided, since denoising, dereverberation, or aggressive segmentation can all remove clinically relevant signal. Even seemingly minor steps, such as voice activity detection thresholds or diarization, can bias measures or introduce irrelevant signal (e.g., an interlocutor’s voice)~\cite{perez2021influence}.

Once constructs and sources of variability are defined, simplicity and transparency should guide measurement. Many voice and speech measures have clear physiological interpretations and can be implemented with established software toolkits. However, discrepancies among extraction tools can lead to inconsistent values of nominally identical measures, complicating cross-study comparisons. What matters, therefore, is not only which measures are extracted, but how: tool choice, parameter settings, frame sizes, and error-handling procedures must be documented so that others can replicate results. Acoustic measures also vary with age and sex, meaning that results must be reported relative to normative distributions. Only by situating measures against appropriate reference values can pathology be distinguished from normal variability.

While recent advances in self-supervised learning and foundational models have produced powerful audio embeddings (\ie vectors that capture salient acoustic information but are not directly explainable), these representations lack direct physiological or clinical interpretability, making it unclear which aspects of the signal they capture and how these relate to underlying clinical constructs. As a result, they may encode confounding factors—such as channel characteristics or language content—as well as sociodemographic biases, potentially limiting their reliability for clinical applications. They should therefore be considered only after interpretable baselines are established, and used to complement clinically meaningful measures to increase trust in model outputs and generalizability. When embeddings are employed, they must be accompanied by bias checks and ablation analyses to ensure that performance is not driven by spurious correlations.

To operationalize the principles of clinically anchored, task-appropriate, and comparable measurement, Section~\ref{sec:feats} introduces a harmonized set of clinically interpretable measures together with normative values for American English, providing a common foundation for the field. These resources are intended for studies designed around specific clinical constructs that manifest in speech—such as phonatory stability, fluency and timing, articulatory precision, or respiratory support -- where acoustic measures serve as valid proxies. They are directly applicable when standardized elicitation tasks are used, namely sustained phonation (e.g. sustaining the vowel \textipa{/A:/}) to probe phonatory control and read speech to capture fluency, prosody, and articulation. The normative values were derived from recordings collected on mobile phones, reflecting both the feasibility of remote data acquisition and the practical conditions under which many clinical studies now operate. For meaningful interpretation, studies should also collect key covariates such as age and sex, and the measures should be used to generate interpretable, reproducible outcomes that can be compared to healthy U.S. English reference norms. The following section details this measure set, the rationale for each measure, and the normative distributions that anchor them.


\section{Measures and Normative Values}
\label{sec:feats}
We describe a set of core measures as a foundation for voice and speech research. We provide detailed descriptions, normative ranges, and open-source code to promote reproducibility, consistency and accessibility across studies. This toolkit serves both as an entry point for new researchers and a benchmark for advanced modeling. 

\subsection{Description of core measures}

\noindent
We define six categories of speech and voice measures: respiration, phonation, articulation, spectral, fluency, and language measures. The language category, which encompasses aspects related to both semantic and syntactic structures, is typically derived from speech transcripts. Although we mention this category here, this paper focuses on measures derived from the acoustic signal. A brief introduction to language measures is given in \Cref{subsec:LanguageProcessing}; for additional information we refer the interested reader  to \cite{lowtext, boschi2017connected, voleti2019review}. Alternative measure sets  and data representation approaches, including those derived from deep learning or foundation models are discussed in \Cref{subsec:featex} and \Cref{subsec:Learnablefeatures} respectively. 

\vspace{\baselineskip}

\noindent 
\textbf{Respiration.} Voice and speech sounds are a direct output of air pressure, carefully and precisely generated by the respiratory system \cite{baken1987clinical}. In the source-filter model of speech production, respiration is considered the power source for voice production; producing louder speech or longer utterances requires a larger inhalation. The use of the respiratory system during speech is known as \textit{speech breathing} \cite{von1982some}. Speech requires precise neuromuscular control of respiratory processes to regulate subglottal pressure (\ie the air pressure below the vocal folds that drives their oscillation) and glottal airflow (\ie the volume of air passing through the vocal folds per unit time during voicing). Inappropriate regulation of subglottal pressure can impair speech intensity and cause unexpected shifts in fundamental frequency. To capture these clinical variations, the respiration measures typically used in the literature, and extracted in this study, primarily relate to the intensity (loudness) of the speech signal and the rate of speech respiration, as detailed in Table \ref{tab:features_respiration}. 


\scriptsize
\renewcommand{\arraystretch}{1.5}

\begin{longtable}{L{2.5cm} L{7.0cm} C{0.8cm} C{1.8cm}}

\caption{Respiratory measures for clinical voice and speech assessment. Features below the dashed line are not in the core set}
\label{tab:features_respiration} \\

\toprule 
\textbf{Measure} & \textbf{Description} & \textbf{Unit} & \textbf{\shortstack{Norm.\\Provided}} \\
\midrule
\endfirsthead

\toprule 
\textbf{Measure} & \textbf{Description} & \textbf{Unit} & \textbf{\shortstack{Norm.\\Provided}} \\
\midrule
\endhead

\midrule \multicolumn{4}{r}{\scriptsize Continued on next page} \\
\midrule
\endfoot

\bottomrule
\endlastfoot

Intensity 
& Sum of the squares of the signal amplitude \citep{boersma2001praat}\textsuperscript{a} 
& dB 
& Yes \\

Intensity Range
& Range of loudness values in a speech signal \citep{boersma2001praat}
& -- 
& Yes \\

\cdashline{1-4}

Voice Range Profile (VRP) 
& Minimum and maximum intensity across set number of frequencies \citep{rychel2023voice} 
& dBHz$^{-1}$ 
& No \\

Number of Breath Events 
& Number of inhalations in a recording \citep{fuchs2021respiratory} 
& -- 
& No \\

Speech Respiration Rate  
& Respiratory rate during speech \citep{fuchs2021respiratory}  
& unit$^{-1}$ 
& No \\

Speech Tidal Volume 
& Amount of air inhaled during a normal breath for speech \citep{winkworth1995breathing} 
& mL 
& No \\

Pause Intervals per Respiration 
& Breathing periodicity \citep{hlavnivcka2017automated} 
& -- 
& No \\

Relative Loudness of Respiration  
& Audibility of respiration relative to loudness of speech \citep{hlavnivcka2017automated} 
& -- 
& No \\    

Respiratory Exchange Latency 
& Time between expiration and respective inspiration \citep{hlavnivcka2017automated} 
& s 
& No \\

\midrule
\multicolumn{4}{L{12cm}}{\footnotesize \textsuperscript{a} This corresponds to the loudness of a speech signal \citep{5312370, itsp2022}.} \\

\end{longtable}

\normalsize
\renewcommand{\arraystretch}{1}

Intensity measures must be interpreted with caution \cite{vsvec2018tutorial}, as they are a composite measure of vocal loudness, reflecting not only source-level factors such as subglottal pressure and the speed of glottal closure, but also harmonic-to-formant resonance interactions and other filter effects. These measures can also conflate true vocal effort with acoustic factors such as microphone distance, room acoustics, and the frequency-dependent directivity of voice radiation. 

Measures within the respiration grouping have potential to serve as \textit{cardio-respiratory acoustic} biomarkers~\cite{Pizzimenti2026}, as they pertaining to respiration and cardiac function. Highlighting this, a range of conditions, including respiratory, neurological, neurodegenerative, and psychiatric disorders, have been shown to disrupt speech breathing mechanisms. Subglottal pressure, in particular, is highly sensitive to changes in neuromuscular control, vocal tract impedance, and ventilatory disorders \cite{baken1987clinical}. Respiratory kinematics are therefore broadly useful for clinical phenotyping: measures associated with rib cage motion, such as speech respiration rate, the loudness of respiration, and respiratory exchange latency, have been shown to differentiate between atypical Parkinsonian syndromes, including progressive supranuclear palsy and multiple system atrophy \cite{hlavnicka2017dysprosody}. Respiratory-driven acoustic limit measures, such as the voice range profile (VRP), are notably reduced across different types of dysphonia and progressively increase following effective treatment \cite{ikeda1999quantitative}. 

Parkinson's disease (PD) offers a clear example of this sensitivity: individuals with PD utilize fundamentally different respiratory strategies than healthy individuals to support speech loudness \cite{sadagopan2007effects, huber2003respiratory}. During speech breathing, speakers with PD often exhibit smaller rib cage volumes and larger abdominal volumes at the initiation of breath groups, demonstrating a more restricted intensity range when attempting to produce sentences across different loudness targets \cite{clark2014loudness}. Respiratory measures have also shown value in machine learning model predicting if patients acute decompensated heart failure are in an admission and discharge states \cite{Riehle2026}.

Normative values are provided only for intensity and intensity range, as these are directly observable acoustic measures. Other measures in this set depend on additional signals or modeling, introducing further uncertainty in their estimation. Following Occam's razor, our \textit{core} normative set prioritizes measures that are transparent and robust. 

\vspace{\baselineskip}

\noindent 
\textbf{Phonation.} Phonation is the process of producing sound through the vibration of the vocal folds. It occurs when air from the lungs passes through the adducted vocal folds, causing them to oscillate. The perceived pitch and quality of the resulting sound depend on the tension of the vocal folds, their rate of vibration, and the applied subglottal pressure. Eventually, phonation reflects the complex interaction between the subglottal pressure and the biomechanical characteristics of the vocal tract, which are shaped by the laryngeal structure and muscle forces arising from reflexive, affective, and learned behaviors \cite{baken1987clinical}.


\scriptsize
\renewcommand{\arraystretch}{1.5}

\begin{longtable}{L{2.5cm} L{7.0cm} C{0.8cm} C{1.8cm}}

\caption{Phonation measures for clinical voice and speech assessment.}
\label{tab:features_phonation} \\

\toprule 
\textbf{Measure} & \textbf{Description} & \textbf{Unit} & \textbf{\shortstack{Norm.\\Provided}} \\
\midrule
\endfirsthead

\toprule 
\textbf{Measure} & \textbf{Description} & \textbf{Unit} & \textbf{\shortstack{Norm.\\Provided}} \\
\midrule
\endhead

\midrule \multicolumn{4}{r}{\scriptsize Continued on next page} \\
\midrule
\endfoot

\bottomrule
\endlastfoot

Fundamental Frequency (F0) 
& Rate of vocal-fold vibration\textsuperscript{a} \citep{baken1987clinical} 
& Hz 
& Yes \\

Pitch Sigma 
& F0 standard deviation, expressed in semitones \citep{baken1987clinical} 
& Semitones 
& Yes \\

Harmonic-to-Noise Ratio 
& Ratio of harmonic to inharmonic spectral energy in voiced speech \citep{boersma2001praat, elemetrics1993multi}
& dB 
& Yes \\

Spectral Slope 
& Slope of the long-term average spectrum \citep{maryn2010acoustic, maryn2010toward} 
& dB/octave 
& Yes \\

Spectral Tilt 
& Tilt of the regression line through the long-term average spectrum \citep{maryn2010acoustic, maryn2010toward} 
& -- 
& Yes \\

Cepstral Peak Prominence 
& Measure of aperiodicity and spectral variation \citep{fraile2014cepstral, murton2020cepstral} 
& dB 
& Yes \\

\cdashline{1-4}

Jitter (Absolute) 
& Average absolute difference between consecutive F0 periods \citep{boersma2001praat, elemetrics1993multi} 
& s 
& No \\

Jitter (Relative) 
& Absolute jitter divided by the average F0 period \citep{boersma2001praat, elemetrics1993multi} 
& \% 
& No \\

Shimmer (local) 
& Average absolute difference between amplitudes of consecutive F0 periods, divided by the average amplitude \citep{boersma2001praat, elemetrics1993multi} 
& \% 
& No \\

Shimmer (dB) 
& Absolute difference between amplitudes of consecutive F0 periods expressed in dB \citep{boersma2001praat, elemetrics1993multi} 
& dB 
& No \\

Percentage of Unvoiced Frames 
& Fraction of pitch frames analysed as unvoiced \citep{boersma2001praat, elemetrics1993multi} 
& \% 
& No \\

Number of Voice Breaks 
& Number of interruptions in the fundamental period during sustained phonation \citep{boersma2001praat, elemetrics1993multi} 
& -- 
& No \\

Degree of Voice Breaks 
& Total duration of voice breaks relative to signal duration \citep{boersma2001praat, elemetrics1993multi}  
& \% 
& No \\

Hammarberg Index 
& Difference between dominant spectral peaks in 0--2 kHz and 2--5 kHz bands \citep{hammarberg1980perceptual}  
& Hz 
& No \\

H1--H2 
& Difference between the first two harmonic amplitudes \citep{hanson1999glottal} 
& dB 
& No \\

\midrule
\multicolumn{4}{L{12cm}}{\textit{Features below require glottal flow estimation; e.g., \citep{drugman2012comparative}.}} \\

H1$^{*}$--H2$^{*}$ 
& Difference between first two harmonics after removing formant influence \citep{hanson1999glottal} 
& dB 
& No \\

Harmonic Richness Factor 
& Amplitude relationship between fundamental and higher harmonics \citep{childers1991vocal} 
& dB 
& No \\

Parabolic Spectral Parameter 
& Quantification of spectral decay of the voice source \citep{alku1997parabolic} 
& -- 
& No \\

Open Quotient 
& Ratio of the open phase to the fundamental period \citep{doval2006spectrum} 
& -- 
& No \\

Closing Quotient 
& Ratio of the closing phase to the fundamental period \citep{doval2006spectrum} 
& -- 
& No \\

Speed Quotient 
& Ratio between opening and closing phases \citep{doval2006spectrum} 
& -- 
& No \\

Normalized Amplitude Quotient  
& Ratio of AC flow amplitude to negative peak of flow derivative, normalised by period \citep{alku2002normalized}  
& -- 
& No \\   

\midrule
\multicolumn{4}{L{12cm}}{\footnotesize \textsuperscript{a} Corresponds to the auditory perception of pitch.} \\

\end{longtable}

\normalsize
\renewcommand{\arraystretch}{1}

Subjective perceptual evaluations play a central role in the clinical assessment of voice quality. For example the the GRBAS scale, which rates five perceptual dimensions: grade of dysphony, roughness, breathiness, asthenia and strain~\cite{hirano1986clinical, nemr2012grbas}. To complement these perceptual assessments with objective metrics, this study describes a comprehensive set of acoustic features related to phonation, as detailed in Table \ref{tab:features_phonation}. A primary acoustic feature in this group is the fundamental frequency (F0), which represents the rate of vocal fold vibration and corresponds directly to the perceptual attribute of pitch. F0 has served as a crucial etiological and symptomatic marker in vocal disorders, though it is important to note that F0, intensity, and spectral properties interact in a highly complex manner to influence overall pitch perception \cite{baken1987clinical}.

Beyond F0, many measures in this category act as markers of overall voice quality and dysphonia severity (\ie impaired voice quality, pitch, or loudness) \cite{maryn2009acoustic, wuyts2000dysphonia, fraile2014cepstral}. 
Although voice quality is a complex perceptual construct driven by physiological changes independent of pitch and loudness \cite{barsties2015assessment, gerratt2001toward}, it can still be objectively measured. For instance, the acoustic measures successfully capture nonmodal or irregular phonations, such as breathiness (\ie audible air leakage), creakiness (\ie low-pitched, irregular vibration), and harshness (\ie rough, strained quality). 

Phonation measures have potential to qualify as \textit{voice} biomarkers ~\cite{Pizzimenti2026}, with measures in Table \ref{tab:features_phonation} mapped directly onto established clinical constructs. For instance, Cepstral Peak Prominence (CPP) reliably distinguishes speakers with and without voice disorders and correlates strongly with perceptual ratings of dysphonia severity \cite{murton2020cepstral}. Closing quotient measures can distinguish dysphonic from non-dysphonic speakers and identify subcategories of mutational dysphonia \cite{lim2007clinical, schoentgen1985acoustic}. Voice breaks can differentiate dysphonia subtypes and relate to vocal effort and self-perceived voice quality \cite{roy2008differential, castillo2023tracking}. 

We selected F0, Pitch Sigma, Spectral Slope and Tilt, and CPP for evaluation in our \textit{core} set due to their robustness to remote data collection and their relative agnosticism to elicitation task (i.e., they remain well-defined in free speech). Measures relying on glottal flow estimation or exhibiting sensitivity to elicitation conditions and/or channel effects are excluded.\\

\vspace{\baselineskip}

\noindent 
\textbf{Articulation.} 
While the vocal source signal originates from the larynx, it is the movement of articulators (lips, tongue, jaw, velum) that shapes the vocal tract resonances into recognizable speech \cite{gillam2024communication}. Acting as a tube resonator, the vocal tract amplifies specific frequencies known as \textit{formants}. By moving the articulators, a speaker alters the shape of the vocal tract to produce distinct phonetic sounds, which can be acoustically characterized by these shifting formant distributions \cite{baken1987clinical}.

\scriptsize
\renewcommand{\arraystretch}{1.5}

\begin{longtable}{L{2.5cm} L{7.0cm} C{0.8cm} C{1.8cm}}

\caption{Articulation measures for clinical voice and speech assessment.}
\label{tab:features_articulation} \\

\toprule 
\textbf{Measure} & \textbf{Description} & \textbf{Unit} & \textbf{\shortstack{Norm.\\Provided}} \\
\midrule
\endfirsthead

\toprule 
\textbf{Measure} & \textbf{Description} & \textbf{Unit} & \textbf{\shortstack{Norm.\\Provided}} \\
\midrule
\endhead

\midrule \multicolumn{4}{r}{\scriptsize Continued on next page} \\
\midrule
\endfoot

\bottomrule
\endlastfoot

Formant Frequencies 
& Centre frequencies of vocal tract resonance peaks \citep{kent2018static} 
& Hz 
& Yes \\

Formant Bandwidths 
& Width of the frequency band 3~dB below the corresponding resonance peak \citep{kent2018static} 
& Hz 
& Yes \\

\cdashline{1-4}

Formant Slopes 
& Change in formant values between two time periods \citep{sandoval2019average}   
& Hz/ms 
& No \\     

Vocal Tract Coordination  
& Cross-correlation between formant frequencies at predefined time delays \citep{quatieri2020noninvasive} 
& -- 
& No \\

Vowel Space Area 
& Area of the quadrilateral formed by the four corner vowels in the F1--F2 space \citep{sandoval2013automatic} 
& -- 
& No \\

Formant Centralization Ratio (FCR) 
& Relationship between F1 and F2 of corner vowels /a/, /u/, and /i/: 
$FCR = (F2_u + F2_a + F1_u + F1_i)/(F2_i + F1_a)$ \citep{sapir2010formant} 
& -- 
& No \\

Vowel Articulation Index (VAI) 
& Reciprocal of FCR: 
$VAI = (F2_i + F1_a)/(F2_u + F2_a + F1_u + F1_i)$ \citep{sapir2011acoustic} 
& -- 
& No \\

Goodness of Pronunciation 
& Posterior probabilities derived from an acoustic model \citep{witt2000phone}  
& --  
& No \\

\end{longtable}

\normalsize
\renewcommand{\arraystretch}{1}

These formants are described by frequency, bandwidth \cite{kent2018static}, or temporal trajectory \cite{kent2020acoustic}. However, the anatomical differences between speakers make absolute comparisons challenging \cite{baken1987clinical}. 
To account for this variability, many of the measures outlined in Table \ref{tab:features_articulation} avoid relying on absolute frequencies and focus on relative, speaker-normalized formant patterns \cite{quatieri2020noninvasive, caverle2020stability}. Beyond formant analysis, the table also incorporates higher-level measures designed to quantify broader pronunciation accuracy and speech intelligibility \cite{witt2000phone}.We selected formant frequencies and bandwidths as our \textit{core} articulation measures. They directly characterize vocal tract resonances and can be estimated from the acoustic signal without additional modelling.

These approaches have direct clinical applications in assessing speech production and speech motor control deficits and are potential speech/articulatory biomarkers~\cite{Pizzimenti2026}. For example, specific measures such as formant centralization, reduced vowel space area (VSA), and reduced second-formant slope, effectively capture the vowel deficits associted with dysarthria \cite{lansford2014vowel}. The goodness of pronunciation measures \cite{witt2000phone}, and its variants have been repurposed to assess both consonant and vowel production precision in pathological speakers, providing objective severity markers for hypernasality, cleft palate, and voice disorders \cite{liu2019acoustical, mathad2021deep, mathad2022consonant}. 

Despite being based on direct measures of articulation, formant based features also have the potential to serve as cognitive/language biomarkers~\cite{Pizzimenti2026}. For example, changes in formant dynamics, captured through vocal tract co-ordination features have been associated with preclinical mild traumatic brain injury \cite{williamson2021using} and expressive language in autism \cite{quatieri2026quantifying} . 

Such overlaps between measures and the clinical constructs they may capture, are common within speech analysis. This makes it essential to work within clearly bounded study designs with well-defined constructs of interest and elicitation tasks, to help avoid ambiguity in what a measure is capturing beyond its strict definition. 

\vspace{\baselineskip}

\noindent 
\textbf{Spectral Analysis.} 
Speech is a highly dynamic signal, with phones changing roughly every 10-100 ms. To capture such variations, speech signals' energy distribution is typically represented by spectrogram using short-time fast Fourier Transform (SFFT) \cite{5312308, itsp2022}. There is always a tradeoff between window size used in SFFT and time-frequency resolutions \cite{5312308, itsp2022}. For example, \textit{wideband} spectral analysis utilises time window between 5 to 10 ms to highlight amplitude variations within each pitch period; \textit{narrowband} analysis utilizes a window size between 20 to 40 ms for clearer frequency resolution of harmonics (\ie, multiples of F0). 

\scriptsize
\renewcommand{\arraystretch}{1.5}

\begin{longtable}{p{2.5cm} p{7.0cm} p{0.8cm} p{1.8cm}}
\caption{Suggested voice and speech measures for clinical health assessment: Spectral.} \label{tab:features_spectral}\\

\toprule 
\textbf{Measure} & \textbf{Description} & \textbf{Unit} &\textbf{\shortstack{Norm.\\Provided}} \\
\midrule
\endfirsthead

\toprule 
\textbf{Measure} & \textbf{Description} & \textbf{Unit} &\textbf{\shortstack{Norm.\\Provided}} \\
\midrule
\endhead

\midrule \multicolumn{4}{r}{{\scriptsize Continued on next page}} \\
\midrule
\endfoot

\bottomrule
\endlastfoot

Spectral Gravity 
& The spectral centroid, or center of gravity, of the spectrum  \citep{buder1996formoffa, mandulak2011can} 
& Hz & Yes \\

Spectral Deviation 
& Spread of frequencies around the centroid (2nd central moment of the spectrum) \citep{buder1996formoffa, mandulak2011can} 
& Hz & Yes \\

Spectral Skewness 
& Symmetry of frequencies around the centroid (3rd central moment of the spectrum) \citep{buder1996formoffa, mandulak2011can} 
& Hz & Yes \\

Spectral Kurtosis & The flatness of the spectrum around the centroid (4th central moment of the spectrum) \citep{buder1996formoffa, mandulak2011can} 
& Hz  & Yes \\  
\cdashline{1-4}

Mel Frequency Cepstral Coefficients 
&  a multivariate spectral representation based on the Mel frequency scale \citep{abdul2022mel, tracey2023towards} 
&  -- & No \\

Linear Predictive Cepstral Coefficients  & a set of cepstral coefficients derived through Linear Predictive Coding \citep{rabiner1993fundamentals} 
& -- & No \\

Perceptual Linear Prediction  
& A multivariate spectral representation based on the Bark frequency scale, additionally equal-loudness pre-emphasis weights are used to simulate the sensitivity of hearing \citep{hermansky1990perceptual} 
& -- & No \\

\end{longtable}

\normalsize

Various techniques can be further applied to characterize the extracted spectrograms, as shown in Table \ref{tab:features_spectral}. 
Spectral Moment Analysis (SMA), for instance, quantifies spectral distributions using different statistical orders: gravity, deviation, skewness and kurtosis \cite{mandulak2011can, buder1996formoffa}. 
While raw FFT representations inherently capture rich information regarding phonation and articulatory muscle control, their high dimensionality makes them computationally heavy.
Different spectral representations are developed to describe the spectral envelope with smaller number of coefficients. This is typically inspired by psychoacoutics research. For example, compact representations include Mel Frequency Cepstral Coefficients (MFCCs) \cite{abdul2022mel, tracey2023towards}, Linear Predictive Cepstral Coefficients (LPCCs), and Perceptual Linear Prediction (PLP) \cite{hermansky1990perceptual, rabiner1993fundamentals}. 

These measures can be used to capture changes in muscle tension and control associated with health conditions and are potential speech/articulatory biomarkers~\cite{Pizzimenti2026}. High-dimensional psychoacoustic representations, in particular, encode linguistic and paralinguistic signals, so could serve as cognitive/language biomarkers as well \cite{Pizzimenti2026}. This possible dual functionality highlights the need to develop collection protocols that tie measures, and ultimately vocal biomarkers, to their elicitation tasks. These representations also capture information about background noise and dataset artifacts and, therefore, should be carefully contextualized when used in digital biomarker applications.

Recently studies have attempted to link these spectral representations to established clinical constructs. For example, the second MFCC coefficient has been linked to existing clinical speech measures, and have been shown to differentiate between healthy and pathological speech groups \cite{tracey2023towards}. Similarly, the spectral mean derived from the long-term average spectrum (LTAS) has proven sensitive to tracking improvements in dysphonia severity \cite{tanner2005spectral}. This growing body of evidence highlights the strong potential of mapping spectral features to robust clinical constructs. 

\vspace{\baselineskip}

\noindent 
\textbf{Fluency.} Fluency refers to the flow flow, continuity, smoothness, rate, and overall effort in speech production. All natural speech contains minor dysfluencies, such as filled pauses (\eg ``uh'' ``uhm'', or ``hm''), unfilled pauses (\eg elongated pauses) or repetitions. Excessive or unusual disruptions can indicate underlying pathology in cognition, respiration, phonation, and articulation. As listed in Table \ref{tab:features_fluency}, fluency measures are generally grouped into three categories: speed, breakdown, and repair. 
Speed measures, such as speaking rate or the mean length of syllables, concern the rate and density of speech delivery. Breakdown measures, such as frequency and duration of pauses, characterize speech interruption. Repair measures, such as hesitations, repetitions, emphasize self-correction. 
\cite{skehan2003task, tavakoli2005strategic}. 

A major advantage of these temporal measures is their computational simplicity. They can often be directly extracted using basic word counts, recording timestamps, and signal energy thresholds. These measures yield robust descriptors that closely correlate with established clinical constructs and measures across health conditions. Fluency measures may potentially serve as either speech-motor or cognitive/language biomarkers \cite{Pizzimenti2026}. Again, which biomarker class a given measure applies to can only be established through validation under a specific protocol and elicitation method.

\scriptsize
\renewcommand{\arraystretch}{1.5}

\begin{longtable}{p{2.5cm} p{7.0cm} p{0.8cm} p{1.8cm}}
\caption{Suggested voice and speech measures for clinical health assessment: Fluency. $^\dag$A spoken unit could be a phoneme, syllable or word.} 
\label{tab:features_fluency}\\

\toprule 
\textbf{Measure} & \textbf{Description} & \textbf{Unit} &\textbf{\shortstack{Norm.\\Provided}} \\
\midrule
\endfirsthead

\toprule 
\textbf{Measure} & \textbf{Description} & \textbf{Unit} &\textbf{\shortstack{Norm.\\Provided}} \\
\midrule
\endhead

\midrule \multicolumn{4}{r}{{\scriptsize Continued on next page}} \\
\midrule
\endfoot

\bottomrule
\endlastfoot

Duration & Total length of a speech recording & sec & Yes \\

Phonation Ratio & Phonation time divided by duration & -- & Yes \\

Speaking Rate & Number of spoken units divided by duration & unit sec$^{-1}$ & Yes \\

Articulation Rate & Number of spoken units divided by phonation time & unit sec$^{-1}$ & Yes \\

Mean Pause Duration  & Mean duration of (filled/silent) pauses & sec & Yes \\

\cdashline{1-4}

Phonation Time & Length of all phonated sounds within the file & sec & No \\

Mean Phrase Duration & Average duration of a phrase (sections of continuous speech
between pauses) & sec & No \\

Coefficient of Variance Phrase Duration & Normalized measure of variability of phrase durations & sec & No \\

Number of Spoken Units & Number of spoken units$^\dag$ identified within a file & -- & No \\

Mean Unit Duration & Phonation Time divided by the number of spoken units & sec & No \\

Mean Length of Run & Average number of units produced in runs of speech between silences & -- & No \\

Number Pauses & Number of (filled/silent) pauses in a recording & unit & No \\

Pause Rate &  Number of (filled/silent) pauses divided by duration  & unit$^{-1}$  & Yes \\

Pause Ratio & Pause Time divided by Recording Duration & -- & No \\

Coefficient of Variance Pause Duration & Normalized measure of variability in the duration of the pauses & -- & No \\

\multicolumn{4}{l}{\textit{Note:} References for the above measures include: \citep{bowden2023systematic, cordella2024connected, cummins2023multilingual, de2021praat, green2004algorithmic, neumann2024multimodal, nevler2019validated, slegers2018connected}} \\ \cdashline{1-4}[0.8pt/4.5pt]
Mean Phone Length & Average duration of  phones~\citep{quatieri2020noninvasive}   &  sec & No \\
Phoneme-Dependent Duration & Linear combination of (subsets) average phone durations~\citep{quatieri2020noninvasive}  &  sec & No \\
Voice Onset Time (VOT) & Time  between the release of a stop consonant and the onset of vocal fold vibration \citep{abramson2017voice} & sec & No \\
Maximum Phonation Time & Maximum length of a continuous phonation of a vowel \citep{maslan2011maximum} & sec & No \\
Pairwise Variability Index & Temporal variability between successive unit (\eg vowel and consonant) intervals \citep{grabe2002durational}  & -- & No \\

\end{longtable}
\normalsize

Variations in these temporal features can be reliable markers for speech production deficits across numerous health conditions, including Huntington's disease, Alzheimer's disease, depression, and schizophrenia \cite{low2020automatedreviewspeech, hecker2022voice, botelho2024speech, ramanarayanan2022speech, neumann2024multimodal, cohen2023multimodal, neumann25_interspeech}. For instance, in ALS-related dysarthria, patients remain consistently slower than healthy controls regardless of their targeted speaking rate \cite{turner1993characteristics}. In Parkinson's disease, speakers exhibit abnormal speech acceleration, with articulation rates negatively correlating with disease duration and intra-word pause rates inversely correlating with clinical motor scores \cite{skodda2008speech}. The voice onset time (VOT) in voiceless syllables have proven effective at quantifying dysphonia severity in patients with vocal hyperfunction \cite{mckenna2020voice}. 

\subsection{Open-source code}
\label{subsec:code}

To demonstrate the extraction of key speech measures aligned with our harmonization framework, we provide an example of how to compute a subset of these measures using the \texttt{senselab}\footnote{https://github.com/sensein/senselab} Python package. This example showcases a straightforward pipeline for extracting clinically relevant speech measures, illustrating both the ease of implementation and the rationale for selecting specific measures based on speech task characteristics and conditions.

Listing~\ref{lst:senselab_example} presents a code snippet that loads raw audio files, preprocesses them by downmixing to mono and resampling to 16kHz for consistency across datasets, and finally extracts health-related speech measurements.

\begin{lstlisting}[style=pythonstyle, caption={Example code for extracting speech measures using \texttt{senselab}.}, label={lst:senselab_example}]
# Install senselab as a dependency
# !pip install senselab

# Import functions and classes
from senselab.audio.tasks.input_output import read_audios
from senselab.audio.tasks.preprocessing import downmix_audios_to_mono, resample_audios
from senselab.audio.workflows.health_measurements.extract_health_measurements import extract_health_measurements

# Load audio files
[audio1, audio2] = read_audios(file_paths=[<audio1_path_here>, <audio2_path_here>])

# Downmix audio samples to mono
audios = downmix_audios_to_mono(audios=[audio1, audio2])

# Resample audio samples to 16kHz (good for measure comparability across different datasets)
audios = resample_audios(audios=audios, resample_rate=16000)

# Extract health-related speech measurements
extract_health_measurements(audios=audios)
\end{lstlisting}

\subsection{Normative values for the core measures}
Establishing normative benchmarks is crucial for clinical applications, enabling the identification of deviations indicative of pathology. For this reason, we computed normative values of our proposed core measures based on speech recordings from the Crowdsourced Language Assessment Corpus (CLAC) \cite{ramani2021clac}. The CLAC dataset consists of audio recordings from diverse speakers performing structured speech tasks (often used in a clinical setting), designed to complement existing clinical speech datasets by providing a healthy reference baseline. For our analysis, we filtered the dataset to include only U.S. participants, ensuring a reference population for English speakers within the United States. The final dataset comprises 1,815 participants, spanning all 50 states and 956 cities, with the largest representation from California, Texas, and Florida. The demographic characteristics are summarized as follows:
\begin{itemize}[nosep]
    \item Gender distribution: 50.5\% (916) females, 49.8\% (903) males, and 0.7\% (13) others.
    \item Age distribution: Mean age of 35.7 years (SD = 11.9), ranging from 18 to 80 years.
    \item Education level: Mean education level of 15.4 years (SD = 2.9), with a median of 16 years.
    \item Presence of speech-related symptoms: 1.4\% (25) participants reported symptoms (e.g., cold, allergies) that could impact speech, while 98.6\% (1,790) did not.
\end{itemize}

Speech measures were extracted from two commonly used clinical tasks: maximum phonation time and the rainbow passage. More details on these tasks are provided in the Supplementary Material and~\cite{ramani2021clac}. These tasks were chosen because they provide structured yet distinct speech conditions: sustained phonation allows for the analysis of vocal stability and phonation quality, while reading introduces natural variations in prosody, articulation, and fluency.

Given the differences between these tasks, measure selection was tailored accordingly. Particularly, measures related to speech timing and fluency (such as pause rate, mean pause duration, speaking rate, and articulation rate) were not extracted for the max phonation task, as they are meaningless for a sustained vowel with no segmental articulation. On the other hand, measures like Harmonics-to-Noise Ratio (HNR), jitter, and shimmer were omitted, as they vary considerably with phonetic and prosodic content, making them unreliable for passage reading.

\begin{table}[!ht]
\centering
\footnotesize
\setlength{\tabcolsep}{3.5pt}
\caption{Normative values for the read passage task. SD = standard deviation; 10th and 90th = 10th and 90th percentiles.}
\label{tab:normative_read}
\adjustbox{max width=\linewidth}{%
\begin{tabular}{l | cccc | cccc | cccc}
\toprule
\textbf{Measure} & \multicolumn{4}{c|}{\textbf{Both Gender}} & \multicolumn{4}{c|}{\textbf{Male}} & \multicolumn{4}{c}{\textbf{Female}} \\
& \textbf{Mean} & \textbf{SD} & \textbf{10th} & \textbf{90th} & \textbf{Mean} & \textbf{SD} & \textbf{10th} & \textbf{90th} & \textbf{Mean} & \textbf{SD} & \textbf{10th} & \textbf{90th} \\
\midrule
\multicolumn{13}{l}{\textbf{Fluency}} \\
\midrule
Duration & 14.7 & 2.9 & 11.9 & 17.9 & 14.6 & 2.6 & 11.7 & 18.0 & 14.8 & 3.2 & 12.0 & 17.7 \\
Phonation ratio & 0.9 & 0.1 & 0.8 & 0.9 & 0.8 & 0.1 & 0.8 & 0.9 & 0.9 & 0.1 & 0.8 & 1.0 \\
Articulation rate & 4.4 & 0.5 & 3.8 & 5.0 & 4.3 & 0.5 & 3.8 & 4.9 & 4.4 & 0.5 & 3.9 & 5.0 \\
Speaking rate & 3.8 & 0.5 & 3.1 & 4.4 & 3.7 & 0.5 & 3.0 & 4.4 & 3.9 & 0.5 & 3.2 & 4.5 \\
Mean pause duration & 0.6 & 0.2 & 0.4 & 0.8 & 0.6 & 0.2 & 0.4 & 0.8 & 0.5 & 0.2 & 0.4 & 0.7 \\
Pause rate & 0.2 & 0.1 & 0.1 & 0.4 & 0.2 & 0.1 & 0.1 & 0.4 & 0.2 & 0.1 & 0.1 & 0.3 \\
\midrule
\multicolumn{13}{l}{\textbf{Respiration}} \\
\midrule
Mean intensity (dB) & 72.7 & 4.8 & 67.0 & 76.3 & 72.1 & 4.8 & 65.7 & 75.9 & 73.3 & 4.7 & 68.2 & 76.5 \\
Intensity SD (dB) & 26.1 & 22.7 & 14.3 & 33.2 & 26.3 & 23.5 & 14.0 & 36.1 & 25.9 & 22.0 & 14.5 & 32.0 \\
Intensity range ratio (dB) & 1.2 & 148.7 & -9.7 & 14.0 & -2.7 & 162.1 & -6.9 & 14.1 & 5.0 & 133.6 & -12.3 & 13.4 \\
\midrule
\multicolumn{13}{l}{\textbf{Phonation}} \\
\midrule
Mean F0 (Hz) & 156.0 & 47.4 & 99.9 & 216.8 & 117.4 & 28.2 & 92.5 & 140.9 & 193.7 & 28.1 & 154.4 & 225.7 \\
F0 SD (Hz) & 32.3 & 17.7 & 14.8 & 51.1 & 23.8 & 18.4 & 12.4 & 32.6 & 40.6 & 12.1 & 26.2 & 56.1 \\
CPP mean & 9.1 & 1.6 & 6.9 & 11.1 & 8.7 & 1.7 & 6.6 & 10.9 & 9.4 & 1.5 & 7.5 & 11.3 \\
CPP SD & 2.2 & 0.5 & 1.6 & 2.9 & 2.2 & 0.5 & 1.5 & 2.9 & 2.3 & 0.5 & 1.7 & 3.0 \\
Mean HNR (dB) & -- & -- & -- & -- & -- & -- & -- & -- & -- & -- & -- & -- \\
HNR SD (dB) & -- & -- & -- & -- & -- & -- & -- & -- & -- & -- & -- & -- \\
Spectral slope & -16.1 & 4.6 & -21.9 & -10.3 & -16.3 & 4.5 & -22.2 & -10.6 & -15.9 & 4.6 & -21.8 & -9.8 \\
Spectral tilt & -0.0 & 0.0 & -0.0 & -0.0 & -0.0 & 0.0 & -0.0 & -0.0 & -0.0 & 0.0 & -0.0 & -0.0 \\
\midrule
\multicolumn{13}{l}{\textbf{Articulation}} \\
\midrule
Mean B1 location & 180.4 & 60.6 & 111.6 & 252.8 & 166.5 & 65.7 & 101.1 & 238.6 & 193.4 & 51.4 & 134.0 & 257.4 \\
Mean B2 location & 380.7 & 120.3 & 232.9 & 542.8 & 314.7 & 91.9 & 209.1 & 441.2 & 445.2 & 109.0 & 313.0 & 583.3 \\
Mean F1 location & 503.6 & 66.4 & 432.1 & 579.4 & 476.6 & 64.7 & 420.3 & 534.3 & 529.8 & 56.8 & 466.5 & 595.0 \\
Mean F2 location & 1557.9 & 128.7 & 1407.1 & 1719.7 & 1512.3 & 98.6 & 1401.3 & 1626.0 & 1602.4 & 138.5 & 1432.1 & 1751.1 \\
B1 location SD & 233.3 & 72.9 & 143.2 & 326.4 & 199.7 & 65.5 & 122.9 & 280.6 & 265.8 & 64.2 & 182.4 & 345.6 \\
B2 location SD & 504.8 & 151.4 & 312.1 & 709.0 & 411.6 & 110.5 & 280.3 & 554.1 & 596.4 & 128.5 & 435.8 & 755.7 \\
F1 location SD & 202.7 & 55.6 & 139.2 & 272.1 & 201.4 & 62.0 & 131.4 & 278.5 & 203.9 & 48.2 & 149.5 & 262.6 \\
F2 location SD & 463.2 & 80.1 & 371.7 & 574.0 & 428.9 & 61.8 & 356.2 & 506.6 & 497.5 & 81.4 & 402.1 & 612.0 \\
\midrule
\multicolumn{13}{l}{\textbf{Spectrality}} \\
\midrule
Spectral gravity & 543.7 & 194.4 & 370.6 & 747.2 & 518.7 & 198.2 & 354.8 & 729.8 & 568.1 & 186.8 & 400.3 & 768.0 \\
Spectral kurtosis & 85.4 & 64.6 & 31.9 & 150.0 & 75.9 & 56.0 & 27.7 & 133.1 & 95.1 & 70.8 & 35.9 & 171.2 \\
Spectral skewness & 4.9 & 1.5 & 3.2 & 6.7 & 4.8 & 1.5 & 3.0 & 6.5 & 5.0 & 1.6 & 3.4 & 6.9 \\
Spectral SD & 397.9 & 139.8 & 241.7 & 585.7 & 395.5 & 138.4 & 241.6 & 588.1 & 399.7 & 140.6 & 241.7 & 576.7 \\
\bottomrule
\end{tabular}%
}
\end{table}

\begin{table}[!ht]
\centering
\footnotesize
\setlength{\tabcolsep}{3.5pt}
\caption{Normative values for the maximum phonation task. SD = standard deviation; 10th and 90th = 10th and 90th percentiles.}
\label{tab:normative_maxphon}
\adjustbox{max width=\linewidth}{%
\begin{tabular}{l | cccc | cccc | cccc}
\toprule
\textbf{Measure} & \multicolumn{4}{c|}{\textbf{Both Gender}} & \multicolumn{4}{c|}{\textbf{Male}} & \multicolumn{4}{c}{\textbf{Female}} \\
& \textbf{Mean} & \textbf{SD} & \textbf{10th} & \textbf{90th} & \textbf{Mean} & \textbf{SD} & \textbf{10th} & \textbf{90th} & \textbf{Mean} & \textbf{SD} & \textbf{10th} & \textbf{90th} \\
\midrule
\multicolumn{13}{l}{\textbf{Fluency}} \\
\midrule
Duration & 16.3 & 7.3 & 8.1 & 26.1 & 17.5 & 7.9 & 8.4 & 27.8 & 15.0 & 6.4 & 7.6 & 23.0 \\
Phonation ratio & 0.9 & 0.2 & 0.5 & 1.0 & 0.9 & 0.2 & 0.5 & 1.0 & 0.8 & 0.2 & 0.5 & 1.0 \\
Articulation rate & -- & -- & -- & -- & -- & -- & -- & -- & -- & -- & -- & -- \\
Speaking rate & -- & -- & -- & -- & -- & -- & -- & -- & -- & -- & -- & -- \\
Mean pause duration & -- & -- & -- & -- & -- & -- & -- & -- & -- & -- & -- & -- \\
Pause rate & -- & -- & -- & -- & -- & -- & -- & -- & -- & -- & -- & -- \\
\midrule
\multicolumn{13}{l}{\textbf{Respiration}} \\
\midrule
Mean intensity (dB) & 70.1 & 11.2 & 60.8 & 77.3 & 69.7 & 7.1 & 60.2 & 76.8 & 70.5 & 14.0 & 61.7 & 77.6 \\
Intensity SD (dB) & 20.0 & 21.3 & 8.7 & 28.3 & 19.2 & 20.7 & 8.4 & 27.6 & 20.8 & 21.8 & 9.2 & 28.4 \\
Intensity range ratio (dB) & 2.9 & 53.8 & -7.7 & 11.7 & 2.2 & 54.1 & -4.7 & 11.2 & 3.7 & 53.3 & -9.7 & 11.8 \\
\midrule
\multicolumn{13}{l}{\textbf{Phonation}} \\
\midrule
Mean F0 (Hz) & 156.6 & 55.5 & 67.0 & 111.6 & 120.2 & 39.3 & 90.3 & 149.4 & 192.6 & 45.0 & 140.9 & 247.4 \\
F0 SD (Hz) & 20.7 & 21.9 & 1.8 & 50.2 & 12.4 & 19.5 & 1.5 & 30.8 & 29.2 & 21.3 & 2.8 & 54.4 \\
CPP mean & 11.5 & 3.0 & 7.7 & 15.4 & 11.8 & 3.3 & 7.5 & 16.0 & 11.2 & 2.6 & 7.7 & 14.6 \\
CPP SD & 1.3 & 1.3 & 0.0 & 3.2 & 1.4 & 1.3 & 0.0 & 3.4 & 1.2 & 1.2 & 0.0 & 3.0 \\
Mean HNR (dB) & 14.0 & 5.2 & 7.2 & 20.6 & 12.0 & 4.8 & 5.4 & 17.5 & 16.1 & 4.8 & 9.9 & 22.2 \\
HNR SD (dB) & 4.0 & 1.2 & 2.6 & 5.6 & 3.8 & 1.1 & 2.6 & 5.3 & 4.2 & 1.3 & 2.6 & 5.9 \\
Spectral slope & -17.9 & 6.9 & -38.6 & -22.6 & -19.5 & 6.2 & -27.4 & -11.9 & -16.3 & 7.2 & -25.1 & -7.0 \\
Spectral tilt & -0.0 & 0.0 & -0.0 & -0.0 & -0.0 & 0.0 & -0.0 & -0.0 & -0.0 & 0.0 & -0.0 & -0.0 \\
\midrule
\multicolumn{13}{l}{\textbf{Articulation}} \\
\midrule
Mean B1 location & 251.1 & 137.8 & 93.6 & 451.9 & 207.8 & 120.9 & 77.9 & 369.3 & 293.0 & 140.0 & 135.3 & 495.2 \\
Mean B2 location & 295.8 & 161.0 & 124.9 & 505.3 & 257.0 & 147.5 & 109.6 & 448.4 & 333.9 & 164.1 & 162.5 & 538.1 \\
Mean F1 location & 714.2 & 135.5 & 556.4 & 879.4 & 670.3 & 113.4 & 537.0 & 803.8 & 757.5 & 142.1 & 573.8 & 922.2 \\
Mean F2 location & 1289.1 & 184.8 & 1074.8 & 1511.7 & 1232.6 & 163.4 & 1055.2 & 1399.6 & 1344.7 & 187.8 & 1115.1 & 1554.2 \\
B1 location SD & 256.5 & 144.1 & 73.1 & 442.9 & 195.1 & 113.6 & 51.5 & 341.0 & 317.2 & 144.8 & 134.5 & 505.6 \\
B2 location SD & 340.1 & 183.6 & 114.2 & 583.8 & 276.1 & 158.4 & 87.0 & 488.7 & 403.5 & 184.1 & 184.7 & 650.5 \\
F1 location SD & 111.4 & 69.0 & 37.2 & 209.3 & 93.0 & 62.8 & 32.1 & 178.0 & 129.5 & 70.1 & 48.8 & 231.8 \\
F2 location SD & 170.5 & 106.6 & 59.1 & 309.7 & 157.8 & 115.6 & 46.1 & 309.9 & 183.1 & 94.9 & 77.2 & 308.7 \\
\midrule
\multicolumn{13}{l}{\textbf{Spectrality}} \\
\midrule
Spectral gravity & 741.0 & 274.1 & 414.6 & 1066.8 & 707.5 & 257.3 & 395.9 & 996.5 & 774.4 & 285.7 & 429.4 & 1129.1 \\
Spectral kurtosis & 30.0 & 50.8 & 5.5 & 58.7 & 32.7 & 48.9 & 6.2 & 62.9 & 27.2 & 52.1 & 4.9 & 49.4 \\
Spectral skewness & 2.7 & 1.9 & 1.0 & 4.8 & 2.9 & 1.8 & 1.2 & 5.1 & 2.6 & 1.9 & 0.9 & 4.4 \\
Spectral SD & 416.4 & 163.8 & 249.1 & 621.2 & 407.2 & 163.5 & 244.4 & 612.2 & 425.4 & 163.9 & 253.6 & 624.8 \\
\bottomrule
\end{tabular}%
}
\end{table}

The normative values in Tables \Cref{tab:normative_maxphon} and
\Cref{tab:normative_read} serve as critical benchmarks for detecting deviations associated with speech impairments, supporting the development of automated speech-based health assessment tools. Below, we analyze the normative values estimated for each category.

\vspace{\baselineskip}
\noindent
\textbf{Respiration.} Variability in intensity provides insights into respiratory control during speech production. Across all participants, the standard deviation of intensity in the max phonation task was approximately 20.0 dB (ranging 19–21 dB for both genders), reflecting the relatively stable loudness expected in a sustained vowel. In contrast, the reading task exhibited a higher intensity standard deviation (approximately 26 dB for both genders). The results are consistent with the natural modulation of speech required for emphasis and phrasing.
While the overall range ratio of intensity was generally higher in phonation (2.9) than in reading (1.2), gender-specific patterns revealed consistent differences across both tasks. Specifically, females exhibited a higher range ratio of intensity than males during both max phonation (3.7 vs. 2.2) and the reading passage (5.0 vs. -2.7). This suggests that females utilize a wider dynamic range of amplitude in both sustained and connected speech. It is important to note that while intensity is informative, it can be sensitive to microphone placement and environmental noise; therefore, these measures should be interpreted with caution.

\vspace{\baselineskip}
\noindent
\textbf{Phonation.} Fundamental frequency (F0) remained consistent across tasks, indicating that speakers sustained phonation within their habitual pitch range. The general average was 156.6 Hz in max phonation and 156.0 Hz in reading, though F0 values naturally clustered by gender. Across tasks,  males averaged ~120.2 Hz and females ~193.7 Hz across tasks. These gender differences correspond to physiological variations in vocal fold length.
Cepstral Peak Prominence (CPP) was consistently higher in phonation (11.5 dB overall) compared to reading (9.1 dB), reflecting a more defined harmonic structure due to the absence of rapid articulation changes. This trend held for both genders, with males showing 11.8 dB and females 11.2 dB in max phonation. Similarly, the mean harmonics-to-noise Ratio (HNR) was 14.0 dB overall in phonation. When split by gender, males exhibited a lower mean HNR (12.0 dB) than females (16.1 dB). HNR was not reported for reading due to its sensitivity to phonetic variability.

\vspace{\baselineskip}
\noindent
\textbf{Articulation.} Vocal tract positioning varied substantially between tasks. In max phonation, the mean first formant (F1) was higher and the second formant (F2) was lower compared to reading, reflecting the more open and stable vocal tract configuration of a sustained vowel. Specifically, F1 averaged 714.2 Hz in phonation versus 503.6 Hz in reading, while F2 averaged 1289.2 Hz in phonation versus 1557.9 Hz in reading. Males consistently showed lower formant values than females across tasks, which correspond to the longer vocal tracts in male speakers \cite{simpson2001dynamic}. For instance, during reading, male F1 was 476.6 Hz (vs. female 529.8 Hz) and male F2 was 1512.3 Hz (vs. female 1602.4 Hz). The standard deviations of F1 and F2 were markedly lower in phonation than in reading for both genders, reinforcing the notion that articulation remains more stable during sustained vowel production.

\vspace{\baselineskip}
\noindent
\textbf{Spectral Analysis.} 
Spectral characteristics demonstrated key differences between tasks. Spectral gravity was higher in phonation (741.0 Hz overall) than in reading (543.7 Hz), indicating  a greater concentration of acoustic energy in higher frequency bands during vowel prolongation. This trend was observed in both male (707.5 Hz vs. 518.7 Hz) and female speakers (774.4 Hz vs. 568.1 Hz).
Spectral kurtosis and skewness were lower in phonation compared to reading, suggesting a smoother spectral profile with fewer sharp peaks. Conversely, spectral standard deviation was slightly higher in phonation (416.4 Hz) than in reading (397.9 Hz), reflecting a broader overall spectral spread in the sustained vowel. In reading, the lower spectral variability is consistent with the averaging of rapidly changing phonetic segments, which produces a less extreme spectral distribution across the passage. When examining this spectral variability by gender, females consistently exhibited slightly higher standard deviations than males in both the max phonation (425.4 Hz vs. 407.2 Hz) and reading tasks (399.7 Hz vs. 395.5 Hz). This increased spectral spread in females is likely driven by their higher F0 and correspondingly wider harmonic spacing. 

\vspace{\baselineskip}
\noindent
\textbf{Fluency.} In the max phonation task, participants sustained phonation for an average of 16.3 seconds. Males sustained phonation slightly longer (17.5 seconds) than females (15.3 seconds). The phonation ratio remained high (0.8-0.9) across groups, reflecting minimal interruptions.
During the reading task, the total duration averaged 14.7 seconds. Females exhibited a slightly faster articulation rate (4.4 vs. 4.3 syllables/second) and speaking rate (3.9 vs. 3.7 syllables/second) compared to males. The overall mean pause duration was 0.6 seconds , though females displayed shorter pauses (0.5s) than males (0.60s). The high phonation ratio in reading (0.9) confirms that the majority of time was spent actively vocalizing, supporting a continuous delivery of the passage.

\section{Discussion and Future Outlook}

Reproducible voice and speech biomarker research is a critical step toward clinical adoption. This work contributes to this harmonization by outlining a set of speech measures, supplying code for their extraction, and providing normative values in read speech and sustained vowels. This contribution is a core step toward the development of best practices and considerations for data acquisition and processing, study design, and transparent result reporting. While the focus of this paper has been on digital biomarker development, this code and the broader harmonization effort are also applicable to other areas of speech-health research. Harmonization is required, for example, in experiments geared toward understanding mechanistic and causal changes in speech and voice. It is also required in early translational research linking speech constructs to health.

Concretely, the code provides a starting point for extracting a bespoke measure space tailored to the elicitation task and clinical outcome. Our plan is that this code will be a living piece of software that is iteratively improved through evidence-based methodological improvements. While we have outlined the most common speech measures utilized across clinical studies and AI applications, rigorous evidence linking these measures to underlying clinical constructs has not yet been fully established. Continued research should focus on rigorously and quantitatively validating these relationships to bridge the gap between exploratory speech analysis and a verified clinical knowledge base.

A core challenge going forward is improving the interpretability of acoustic measures beyond speech science. While some measures may be interpretable to speech specialists, they can be difficult to relate to clinical decision-making or patient experiences. To support translation, this article provides perceptual and physiological correlates of acoustic measures that can help translate measure patterns into more interpretable physiological or perceptual patterns (e.g., low f0 variance and monotonous speech; low HNR to breathy or rough voice). 

As the field evolves, we emphasize the importance of careful design across the voice and speech product lifecyce including of speech elicitation tasks and speech representations, confounder identification, robust evaluation, and comparison with transparent frameworks. Robust and agreed upon standards will provide a foundation for large-scale replication studies and robust validation of speech and voice biomarkers, supporting their clinical integration as predictive health indicators.

\section*{Acknowledgments}
NC is part funded by the National Institute for Health and Care Research (NIHR) Biomedical Research Centre (BRC): Maudsley. The views expressed are those of the authors and not necessarily those of the NHS, the NIHR or the Department of Health and Social Care.
FC currently works for Apple.

\section*{Disclaimer}
Massachusetts Institute of Technology Lincoln Laboratory disclaimer: Approved forpublic release. Distribution is unlimited. This material is based upon work supported by the Department of the Air Force under Air Force Contract No. FA8702-15-D-0001 or FA8702-25-D-B002. Any opinions, findings, conclusions or recommendations expressed in this material are those of the author(s) and do not necessarily reflect the views of the Department of the Air Force.


\suppappendix{Study Design.}
\label{subsec:StudyDesign}
This first phase involves defining the project’s scope, goals, and methodology. Key elements include:
\begin{itemize}

\item \textit{Target Condition Selection}: Specify the health condition(s) targeted for analysis, as well as the task to be solved, or the research question to be answered. 

\item  \textit{Identification of Clinical Constructs and Symptoms}: Define the clinical constructs (e.g., cognitive impairment, emotional dysregulation) that are core to the target condition, especially those likely to have observable speech manifestations. This includes identifying specific signs or symptoms within these constructs that can be reliably measured through speech (e.g., language complexity in cognitive impairment, vocal tone in emotional dysregulation). This step bridges clinical understanding with observable speech-based markers.

\item \textit{Hypothesis/Research Questions Framing}: In hypothesis-driven studies, preliminary hypotheses may be proposed to describe expected speech manifestations of the condition (e.g., distinctive vocal or language measures). For exploratory studies, where predefined hypotheses are not required, this step may instead involve outlining broader domains of interest or open research questions.

\item \textit{Participant Inclusion and Exclusion Criteria}: Establish criteria for participant selection, accounting for demographics (e.g., age, gender), language background, and other relevant factors to ensure a representative sample and minimize variability unrelated to the condition.

\item \textit{Elicitation Methods and Collection Protocol}: Define methods for speech data collection, detailing elicitation tasks (e.g., reading passages, spontaneous speech), recording set-up, session length, and conditions. The selection of speech elicitation tasks in particular plays a critical role in linking speech measures to specific clinical constructs; this is discuss further in~\ref{subsec:elecit}.

\item \textit{Definition of Success Evaluation Criteria}: Establish measurable criteria for assessing the success of the project. This may include evaluating data quality (e.g., speech recording clarity, adherence to protocols), identifying existing datasets or benchmarks for comparison and parallel validation, and establishing task-specific performance measures that are appropriate for the task.
\end{itemize}

\noindent
Because voice and speech are highly sensitive to methodological choices, we recommend pre-registration of study protocols; \eg \cite{cummins2025protocol}. Pre-registration clarifies hypotheses, analytic strategies, and outcome definitions, thereby reducing risks of bias and selective reporting. Publicly available protocols foster reproducibility and harmonization across the community.

\vspace{6pt}
\noindent
\textbf{Existing Benchmarks}
The research community has organized multiple challenges in the area of speech biomarkers, which have introduce or curate new datasets, providing common test beds for researchers to benchmark and compare their approaches. Among the most renowned challenges are the ComParE and AVEC challenges.
ComParE (Computational Paralinguistics Challenge) is an annual challenge, launched in 2009, which has addressed various health-related tasks, including as sleepiness~\cite{schuller2011interspeech}, intoxication~\cite{schuller2011interspeech}, Parkinson’s disease~\cite{schuller2015interspeech}, cold~\cite{schuller2017interspeech}, and COVID-19~\cite{schuller2019interspeech}.
AVEC (Audio/Visual Emotion Challenge) focuses on multimodal analysis of emotional and psychological states, including depression~\cite{Valstar2016avec, ringeval2019avec} and bipolar disorder~\cite{ringeval2018avec}.

More recently, several challenges have focused exclusively on detecting Alzheimer’s disease, including ADReSS~\cite{ADReSS}, ADReSS-o~\cite{luz2021adress_o}, ADReSS-M~\cite{luz2024overview} (curated subsets of the Pitt Corpus~\cite{becker1994natural}), Taukadial~\cite{garcia2024connected}, and the PROCESS challenge\footnote{\url{https://processchallenge.github.io/}}.

Despite their utility, these benchmarks present limitations. For example, both the Pitt Corpus and the C19C corpus used for COVID-19 detection have been criticized for bias.  The Pitt Corpus contain dataset- and interviewer-related confounds that can influence automated Alzheimer’s disease assessment \cite{perez2021influence,liu2024clever}, while the C19C corpus contains recording-related biases that may lead to spurious COVID-19 predictions \cite{solera2021transfer}. Consequently, while these datasets enable comparative evaluations, results should be interpreted with caution. Whenever possible, new data should be collected in accordance with the guidelines outlined in this document.

\suppappendix{Speech Elicitation Considerations.}
\label{subsec:elecit}

The selection of speech elicitation tasks plays a central role in linking speech measures to specific clinical constructs. Different voice and speech elicitation tasks engage distinct cognitive, linguistic, and motor processes~\cite{ramanarayanan2022speech}. 

Therefore, to develop accurate voice- or speech-based health measures, tasks should be intentionally selected to amplify or reveal symptoms relevant to the condition under study; \eg cognitive impairment, affective dysregulation, or speech-motor dysfunction \cite{ramanarayanan2022speech}. For example, diadochokinetic exercises and reading tasks can be used to highlight speech-motor deficits~\cite{ramanarayanan2022speech}, while naming and picture description tasks can be used to assess neurocognitive impairments~\cite{boschi2017connected}. Selecting tasks that place appropriate cognitive or sensorimotor demands on the speaker allows researchers to capture signal variations that correlate with the severity of the health condition, thereby improving the sensitivity and specificity of the derived speech health measures. Failure to carefully design these tasks can lead to the collection of speech that lacks the necessary variance to reveal clinically relevant measures.

Additional considerations when selecting tasks include:
\begin{itemize}
    \item The use of a warm-up and familiarization task for your voice
    \item Potential practice effects and their impact on performance
    \item Task ordering and the possibility of bias introduction
    \item Task acceptability (ideally established through participant feedback or co-design)
    \item Collection procedure (some tasks, e.g., sustained phonation, may require more explicit instructions than others, which could influence sample validity). 
    \item Workload associated with data preparation, such as manual segmentation, transcription, or diarization. While automated tools can reduce this burden, they often come with their own limitations that require careful validation.
\end{itemize}

\noindent
\textbf{Example Tasks.} The choice of speech elicitation tasks should be guided by the target condition, as different tasks emphasize distinct symptomatic measures. Commonly used tasks include:
\begin{itemize}
    \item Diadochokinetic (DDK) tasks -- Rapid repetition of syllables (e.g., "pa-ta-ka"). 
    \item Sustained phonation or maximum phonation time (MPT) -- Sustaining a vowel sound on a single deep breath for as long as possible at a comfortable pitch. Variants involve different vowels or pitch modulations.
    \item Verbal fluency tasks -- Listing words within a semantic category, e.g., animals, food names (semantic fluency) or starting with a given letter (phonemic fluency). See further guidelines in~\cite{spreen1998compendium}.
    \item Picture description tasks -- Describing an image, often the Cookie Theft picture~\cite{Goodglass2001}.
    \item Reading tasks -- Reading standardized passages such as the Rainbow Passage, the Grandfather Passage, or Aesop's fable ``The North Wind and the Sun''.
    \item Monologues -- For example, describing a daily activity, or a memory.
    \item Dialogues -- For example, semi-structured interviews to elicit spontaneous speech.
\end{itemize}

\noindent
Numerous other tasks exist, systematically categorized in the literature. For instance, Ng et al.~\cite{ng2024tutorial} describe speech tasks based on response freedom and computational or sensorimotor demands inherent to completing a task. Low et al.~\cite{low2020automatedreviewspeech} discuss the advantages of each task to support appropriate task selection.

\suppappendix{Data Collection.}
\label{subsec:DataCollect}
Recording equipment and conditions directly affect the capture of voice and speech data. Therefore, careful consideration and detailed reporting of factors related to this aspect of the pipeline are crucial for reliably identifying and validating voice and speech biomarkers. During recording, vocal sounds are captured to produce digital speech data; the quality and consistency of this process are crucial for reliable analyses and valid assessments of health outcomes. 

Measurement errors and biases often arise from variations in multiple factors, including
\begin{itemize}
    \item \textit{Recording Devices}: microphone types (e.g., professional vs. mobile), device cost and accessibility, device-specific frequency responses and signal processing \cite{awan2024evidence}.
    \item \textit{Recording Setups and Environments}: microphone distance, background noise, room acoustics.
    \item \textit{Digitization}: sampling rate, bit depth, compression codec.
    \item \textit{Speaker Characteristics and Sociodemographic Factors}: gender, sex, age, education level, primary language, clinical conditions.
    \item \textit{Participant Behavior}: adherence to instructions, vocal behaviors (e.g., voice intensity), effort.
    \item \textit{Researcher Behavior} differences in task administration, feedback (e.g., encouragement), or tasking.
\end{itemize}

\noindent
To reduce downstream biases, harmonizing recording protocols across participant groups is essential. This includes using consistent devices and microphones, standardizing recording environments, and providing clear, uniform instructions to encourage natural speech. In remote or unsupervised settings, gathering details about the recording environment and conducting pilot studies can identify issues with adherence and variable conditions. Addressing these factors enhances data quality, reliability, and the generalizability of findings across diverse populations.

Beyond audio recording factors, collecting metadata on additional speaker characteristics is essential to account for potential confounders of health status. These can include sex, age, languages spoken, education level, presence of specific learning difficulties (SpLD), occupation, daily voice use, accent or nativeness, speech-related symptoms (e.g., speech, language, psychiatric, neurological, movement), medications \cite{fusaroli2023identifying}, and general health or comorbidities (e.g., fatigue, heavy smoking, prior otolaryngology or dental surgery, current alcohol levels, hydration, menstruation status). Incorporating this information alongside speech measures during analysis can improve health status prediction by adjusting for confounding factors and assessing their impact. 

\suppappendix{Pre-processing Considerations.}
\label{subsec:preprocess}
Pre-processing prepares raw digital voice and speech data for analysis and can include normalization, denoising, dereverberation, speech enhancement, speaker diarization, speaker separation, and automatic speech recognition (ASR) ~\cite{mehrish2023review}. 

\begin{itemize}
    \item \textit{Normalization}: aligning volume across recordings reduces unintended variability. Such adjustments could suppress specific measures related to loudness that are reflective of certain health conditions.
    \item \textit{Denoising/Dereverberation}: techniques that remove background noise and room echo from the audio signal. While they improve perceptual clarity, they can also distort spectral measures, obscuring biomarker-relevant information \cite{zhang2018deep}.
    \item \textit{Diarization}: the process of segmenting an audio recording into speaker-specific regions. This is critical for isolating participant speech from interviewer speech, but is error-prone when the number of speakers is unknown.
    \item \textit{Automatic Speech Recognition (ASR)}: converting speech audio into text using trained models. State-of-the-art systems still struggle with disordered or atypical speech and may filter out disfluencies that are clinically informative \cite{botelho2024tackling}. Generative AI–based ASR can also introduce hallucinations \cite{baranski2025investigation}.
\end{itemize}

\noindent
The choice of methods influences the accuracy, reliability, and clinical relevance of any resulting voice or speech biomarkers. Each pre-processing step introduces uncertainty and may propagate error. Minimizing the use of such tools, while reporting performance both with and without pre-processing, helps ensure transparency.

\suppappendix{Feature Extraction.}
\label{subsec:featex}
Measure extraction converts voice and speech signals into structured representations that capture their temporal, spectral, and prosodic characteristics. These representations serve as the foundation for computational modeling of speech-based behaviors. Several software libraries are widely used in research for this purpose, each offering distinct analytical advantages:
\begin{itemize}
    \item \textit{Praat} \cite{boersma2001praat} and its Python interface \textit{parselmouth} \cite{jadoul2018introducing} are standard tools in phonetic research, enabling precise extraction of articulatory and phonatory measures such as pitch, intensity, formants, and voice quality measures.
    \item \textit{OpenSMILE} \cite{eyben2010opensmile} is a Python package for large-scale acoustic feature extraction, including standardized configurations such as eGeMAPS \cite{eyben2015geneva} and ComParE \cite{weninger2013acoustics}, which have become benchmarks in paralinguistic and affective computing challenges.
    \item \textit{Librosa} \cite{mcfee2015librosa} is a Python package originally developed for music information retrieval. It offers versatile implementations of spectral and cepstral representations (e.g., MFCCs, chroma measures, and mel-spectrograms), making it widely adopted for audio signal analysis in both speech and non-speech contexts. In most cases, measures are extracted through a probabilistic approach.
    \item \textit{Torchaudio} \cite{yang2022torchaudio} is a Python package for GPU-accelerated preprocessing, transformations, and feature extraction (e.g., log-mel spectrograms, MFCCs), facilitating efficient integration within deep learning pipelines.
\end{itemize}

\noindent
The choice of toolkit and the specification of measures to compute, analyze, and report as clinical measures can lead to variations in the resulting measures. To advance the operationalization of voice and speech measures as digital biomarkers, systematic evaluation across diverse tasks, populations, and conditions is required. Such efforts are critical to determining which measures generalize beyond task-specific contexts and are most suitable for clinical applications. However, many studies continue to rely on pre-defined measure sets developed for narrowly defined objectives (e.g., condition-specific models), which may limit generalizability and hinder cross-domain translation.

A key step toward clinical translation involves the development of \textbf{tailored clinical speech representations}. These should be adapted to the characteristics of the clinical population and experimental context, rather than derived from consumer-oriented speech AI pipelines optimized for generic or commercial applications. Aligning measure design with the elicitation task and mapping extracted measures to relevant clinical constructs can help ensure that analyses target speech markers most indicative of the intended clinical outcomes. Establishing such mappings also mitigates the influence of confounding variables and enhances the interpretability of the results.

\suppappendix{Data Analysis.} 
\label{subsec:DataAnalysis}
This phase entails applying statistical or machine learning methods to analyze and extract insights from the processed audio data. 

\vspace{6pt}
\noindent \textbf{Statistical Methods:} are an essential aspect of clinical research which are used to determine if there is a measurable relationship, difference, or association between variables in a set of data; \eg a speech measure and a clinical outcome. Broadly, the main forms of statistical methods are \textit{descriptive statistics}, which summarize data using indexes such as the mean and median, and \textit{inferential statistics}, which use statistical tests to draw conclusions from data. Conducting appropriate statistical analysis \cite{mishra2019selection} is a crucial step towards establishing the clinical validity and reliability of speech and voice measures. For example, mixed-effects models have been effectively used to account for both fixed effects (e.g., confounds) and random effects (e.g., individual variability) in longitudinal speech biomarker studies \cite{cummins2023multilingual, lewis25_interspeech}, enabling more robust and generalizable conclusions.

\vspace{6pt}
\noindent \textbf{Machine Learning:} is a family of artificial intelligence modeling approaches used to learn patterns from data (and labels) and make \textit{predictions}. These approaches can be broadly categorized into: (i) Classification methods, that assign data to predefined health-related categories; (ii) Regression models, that predict continuous health outcomes, such as disease severity; and (iii) Clustering techniques, that group data points based on intrinsic similarities. Within health applications, the of choice of baseline to compare results against is particularly important as weak comparison against meaningless baselines create over-optimistic expectations \cite{demasi2017meaningless}.   

\vspace{6pt}
The core difference between these methods is that statistical analysis typically focuses on understanding within-sample associations (e.g., hypothesis testing, correlations, or regressions), whereas machine learning emphasizes building models that generalize well to independent, unseen test data, ensuring robust predictive performance beyond the original dataset. Both approaches play important roles in the identification and verification of voice and speech biomarkers.

Statistical analysis highlights associations between single acoustic variables and health states, providing evidence of a potential biomarker. While demonstrating an association is essential, it may not be sufficient for clinical screening if the measure lacks adequate sensitivity or specificity, resulting in excessive false positives or negatives. Therefore, a key validity criterion for a biomarker should be its out-of-sample predictive performance and generalizability, which robust machine learning analyses aim to provide \cite{low2024identifying}. Ensuring that test data remains completely unseen during the modeling process is critical. Given the limited availability of data, researchers often employ cross-validation; however, improper implementation can lead to overly optimistic performance estimates. To prevent information leakage during model selection and hyperparameter tuning, nested cross-validation has to be employed \cite{low2020automatedreviewspeech, ghasemzadeh2024toward}.

\suppappendix{Health State Assessment.}
\label{subsec:HealthAssessment}
This final phase involves an interpretative analysis of the results to evaluate health states, extract clinical insights, and derive tailored intervention plans.

For speech and voice markers to support informed clinical decision-making, models must be both reliable and interpretable. We highlight three key requirements: (1) temporal consistency to ensure stable results over relevant time frames, \ie test-retest reliability; (2) interpretability to foster trust and transparency in model decisions; and (3) real-world implementation considerations, including net benefit analysis and continuous evaluation to accommodate population and technological shifts.

\begin{itemize}
    \item \textit{Test-retest reliability}: To be considered a biomarker, voice and speech measures must produce consistent results over a time interval short enough for the underlying health state to remain stable, so that any variability reflects measurement error rather than genuine clinical change. Evidence shows that achieving clinically meaningful test–retest reliability using voice and speech challenging \cite{stegmann2020repeatability,feng2024test}.
    \item \textit{Interpretability}: This is the ease of understanding how inputs influence a model’s outputs. During biomarker development, it is crucial to confirm that predictions accurately reflect meaningful relationships with the underlying health state, rather than spurious correlations or bias. This fosters clinical trust, supports regulatory approval, and informs the refinement of both models and data collection. Transparent models (e.g., linear models, decision trees) provide direct insight into measure importance, while black-box models require additional methods (e.g., XAI toolkits such as LIME, SHAP) to explain their behavior—especially when inputs are less intuitive, such as spectrograms or deep learning embeddings \cite{akman2024audio}. It is worth noting that XAI does not demonstate that the model learned a clinically meaningful or causal relationship. These methods typically describe model behavior and should not replace confound checks or clinical validation.
    \item \textit{Real-World Implementation}: Since the relationship between voice and speech measures and health outcomes can vary with demographics, disease severity, comorbidities, extraction methods, or recording devices (dataset shift \cite{wiles2021fine}), ongoing evaluation on updated, held-out samples is crucial to maintain accuracy, reliability, and clinical relevance. Deploying a voice- or speech-based biomarker directly influences clinical decisions, from treatment initiation to costly follow-ups. While sensitivity, specificity, and precision are key for initial validation, net benefit analysis \cite{vickers2016net} is essential at later stages to evaluate the real-world impact of decisions based on the biomarker, ensuring it ultimately improves—not compromises—patient care.
\end{itemize}

\suppappendix{Language Processing.} 
\label{subsec:LanguageProcessing}
Processing on transcribed speech offers values for healthcare applications. It helps assess cognitive function, language development, and early signs of neurological or psychological disorders. Measures within this class have potential to serve as Cognitive/Language biomarkers \cite{Pizzimenti2026}. Various natural language processing (NLP) methods have been used to automatically extract speech measures linked to clinical constructs  \cite{lowtext, boschi2017connected, voleti2019review, TOKAREVA2026122488}. For instance, lexicon-based methods count predefined, expert-validated words and phrases. While highly interpretable, they require manual curation, often miss context-dependent expressions (false negatives), and struggle with context reasoning (false positives). Lexical diversity measures (e.g., TTR, Yule’s K, MTLD, HD-D) quantify the variance in a speaker’s vocabulary without deep semantic analysis. TF-IDF provides a data-driven way to discover words associated with dependent variables without expert curation, though it lacks semantic understanding and may not generalize across different datasets.

Alternatively, word embeddings represent language as fixed-dimensional vectors to capture deeper semantic meaning \cite{lowtext}. By computing the cosine similarity between word or utterance embeddings (using models like word2vec, FastText, or GloVe), researchers can objectively measure complex clinical constructs; \eg \cite{perez2025exploring}. For example, similarities between embeddings of adjunct words or sentences were used to measure semantic clusters and switches, speech coherence to assess psychosis, thought disorder and schizophrenia. These similarity measures were also shown to correlate to clinical measures such as Communication Disturbances Index, Thought and Language Disorder (TALD) ratings, and social cognition ratings when assessing the respective diseases \cite{docherty1996communication, xu2020centroid, joulin2016fasttext, tang2023clinical}. 

Generative large language models (LLMs) are state-of-the art for all NLP tasks since they can tasked directly to perform text classification, named-entity recognition, among other tasks \cite{lowtext}. Even if their underlying deep learning architecture is not fully interpretable, generative LLMs can successfully analyze documents to output interpretable categorical scores, such as hesitation, sadness, or disorganized speech—for direct clinical evaluation and provide rationales for their outputs and verbatim quotes \cite{lowtext, botelho2024macro, perez2025exploring}. They can even generate new expert lexicons if interpretable or lightweight models are preferred \cite{low2025using}.

\suppappendix{Learnable Data Representations.} 
\label{subsec:Learnablefeatures}
Audio embeddings are compact, high-dimensional representations of speech signals, typically derived from deep neural networks that transform raw waveforms, or spectral representations, into lower-dimensional vectors suitable for machine learning applications (e.g., recognizing emotions from speech; voice identification). Contemporary embeddings are often derived from foundational models trained with self-supervised learning on large-scale unlabeled audio, optimizing objectives such as predicting masked segments of the input.  Foundational models are often trained on vast amounts of audio data in general, including diverse audio sources, such as speech, music, and environmental noises, enabling them to learn rich general-purpose representations.

Commonly used audio embeddings include: 
\begin{itemize}
    \item \textit{x-vectors}: deep neural network based speaker embeddings, proposed as an alternative to i-vectors for speaker \cite{snyder2028x-vectors} and language recognition \cite{snyder2017deep} tasks. 
    Several enhancements have been introduced, notably the ECAPA-TDNN architecture \cite{desplanques2020ecapa-tdnn}.
    \item \textit{Wav2Vec 2.0}: Developed by Facebook AI Research, this model processes raw audio directly and generates embeddings that capture both local and contextual measures of the audio signal~\cite{baevski2020wav2vec2}.
    \item \textit{Hubert}: A SSL model trained to learn pseudo-phonemes using Cross-Entropy plus k-Means Loss~\cite{hsu2021hubert}.
    \item \textit{TRILLsson}: a series of "all-purpose" paralinguistics models~\cite{shor2022trillsson} that relies on knowledge distillation from the non-public "Conformer Applied to Paralinguistics" (CAP12) model~\cite{shor2022universal}), using publicly available speech data--Libri-light~\cite{kahn2020libri} and AudioSet~\cite{gemmeke2017audio}. 
    \item \textit{VGGish}: Based on the VGG architecture, this model generates embeddings from audio spectrograms and is commonly used for various audio classification tasks~\cite{hershey2017cnn}.
    \item \textit{YAMNet}: A deep learning model that generates embeddings from audio data, particularly useful for environmental sound classification and speech analysis~\cite{plakal2020yamnet}. 
    \item \textit{OpenL3}: This model generates embeddings from audio using a deep neural network trained on a large dataset of audio and video pairs, capturing both audio and visual measure~\cite{arandjelovic2017look, cramer2019look}.
\end{itemize}

\noindent
The field of audio-embedding are heavily dominated by grey literature, and transparent evaluations of clinical data remain rare. This is concerning as, similar to high dimensional spectral representations, audio embeddings encode more than just health-related information, which can easily confound machine learning models. Additionally, foundational models are often proprietary “black boxes” further raising inclusivity and generalisability concerns. To mitigate these risks, these embeddings should be used cautiously, with a detailed analysis and identification of potential biases that may lead to overestimated performance.

Other researchers have leveraged deep learning to extract interpretable dimensions to represent the data, instead of "black-box" audio embeddings. Examples of this line of research include neural approaches that explicitly learn a mapping from a high-dimensional measure layer into an interpretable measure layer as part of the overall deep neural network model~\cite{jiao2017interpretable, leschly2025exploration}. However, while these dimensions themselves are interpretable, the extraction process remains opaque, and thus caution is advised when interpreting the results. Nonetheless, evaluating these interpretable dimensions for clinical validity may offer valuable insights.

\end{document}